\documentstyle[12pt,aaspp4,tighten]{article} 

\def\h{H$_2$}

\def\deg{{$^\circ$}}
\def\sun{$_\odot$}
\def\Msun{M$_\odot$}

\def\Lsun{L$_\odot$}
\def\Msunyr{M$_\odot$\,yr$^{-1}$}
\def\Msunkmsyr{M$_\odot$\,km\,s$^{-1}$yr$^{-1}$}
\def\gtsim{\lower.5ex\hbox{$\buildrel > \over\sim$}}
\def\ltsim{\lower.5ex\hbox{$\buildrel < \over\sim$}}

\slugcomment{To be published in IAU Symposium 227 Proceedings on
  Massive Star Formation: A Crossroads in Astrophysics.}

\begin{document}

\title{Massive Star Outflows}

\author{Debra Shepherd\altaffilmark{1}} 

\vspace{-3mm}
\altaffiltext{1}{National Radio Astronomy Observatory, P.O. Box 0,
Socorro, NM 87801}

\begin{abstract}
Molecular outflows in the form of wide-angle winds and/or
well-collimated jets are associated with young stellar objects of all
luminosities.  Independent studies have established that the mass
outflow rate is proportional to L$_{bol}^{0.6}$ for L$_{bol} = 0.3$ to
$10^5$ L$_{\odot}$, suggesting that there is a strong link between
accretion and outflow for a wide range of source luminosity and there
is reasonable evidence that accretion-related processes are
responsible for generating massive molecular flows from protostars up
to spectral type B0.  Beyond L$_{bol} \sim 10^4$ L$_{\odot}$, O stars
generate powerful wide-angle, ionized winds that can dramatically
affect outflow morphology and even call into question the relationship
between outflow and accretion.

Recently Beuther \& Shepherd (2005) have proposed an evolutionary
scenario in which massive protostellar flows (up to early B spectral
type) begin collimated.  Once the star reaches the Main Sequence,
ionizing radiation may affect the balance between magnetic and plasma
pressure, inducing changes in the flow morphology and energetics.
Here I review the properties of outflows from young OB stars, discuss
implications and observational tests of this proposed evolutionary
scenario, and examine differences between low-mass and massive star
formation.

\end{abstract}


\section{Introduction}

The dynamics of outflow and infall associated with young stellar
objects (YSOs) affect the turbulent support and dissipation of
molecular clouds, the final mass of the central star, and ultimately,
the conditions for planet formation.  Outflows in the form of
wide-angle winds and/or well-collimated jets are associated with YSOs
of all luminosities.  However, the conditions and timescales
associated with massive star formation differ from their low-mass
counterparts, e.g.: 1) Massive stars evolve to the Zero Age Main
Sequence (ZAMS) more rapidly than low-mass stars: O spectral type
stars have a Kelvin-Helmholtz timescale $\ltsim 10^4$ years while
solar-type stars require more than $10^7$ years to reach the main
sequence;
2) OB stars reach the ZAMS while still embedded and perhaps still
accreting.  Once on the ZAMS, UV photons generate a hyper-compact HII
(HC HII) region and strong winds drastically affect the physical
conditions, structure, and chemistry of their surrounding;
3) OB stars form in denser clusters than lower mass stars and a higher
fraction of OB stars form in binary or multiple systems.  Thus, there
may be more dynamical interactions between (proto)stars
(e.g. Bonnell, Bate, \& Vine 2003).

Two scenarios have been proposed to explain the formation of massive O
stars.  The first scenario relies on non-spherical accretion which
over comes radiation pressure even from the most luminous stars (e.g.,
Norberg \& Maeder 2000, Behrend \& Maeder 2001, Yorke \& Sonnhalter
2002, McKee \& Tan 2003).  However, the most massive star that can
form in the 2D models of Yorke \& Sonnhalter is about 30\,M{\sun} --
even when 120\,M{\sun} of gas is available in the initial cloud.  A
solution to this problem was presented by Krumholz et al. (2005).
Using a 3D Monte Carlo radiative transfer code (Whitney et al. 2003)
modified to account for time-dependent hydrodynamics Krumholz et
al. found that the optically thin outflow channels carried away
significant radiation, allowing the surrounding envelope to remain
relatively cooler.  This made it easier for the core to collapse to
form very massive stars.  The wider the outflow cavity, the more
easily radiation can escape.  Thus, the presence of a cavity created
by a bipolar outflow may be critical for the formation of the most
massive stars.  Yet, the accretion-based formation of massive stars
may differ from low-mass scenarios.  In particular, recent simulations
suggest that nearly 90\% of the mass of massive stars is not due to
accretion from an initial clump or envelope but from subsequent
competitive accretion of material much farther from the stars (e.g.,
Bonnell, Vine, \& Bate 2004).

The second scenario proposed for the formation of O stars involves the
coalescence of lower mass (proto)stars at the centers of dense
clusters (e.g., Bonnell, Bate, \& Zinnecker 1998, Stahler, Palla \& Ho
2000, Bonnell \& Bate 2002).  The mergers would likely destroy any
accretion disks around the lower mass components and disrupt their
outflows.  The resulting massive star will likely have rotating
circumstellar material but accretion is not a necessary criteria for
formation.  Simulations by Bonnell, Bate, \& Vine (2003) suggest that,
while true mergers of (proto)stars may be rare outside of binary
systems, dynamical interactions between stars in dense star forming
clusters are not: nearly 1/3 of all stars and most stars with
M$_{\star}> 3$\,M{\sun} suffer disruptive interactions that can
truncate circumstellar disks and bring the accretion and outflow
process to an abrupt and possibly explosive halt (Bally \& Zinnecker 2005).
While the competitive accretion model predicts molecular outflows with
properties that may be similar to low-mass outflows, it is unlikely
that highly-collimated structures could exist in the coalescence
scenario and the outflow energetics may be significantly different.

Previous reviews that focus mostly on outflows from low-mass YSOs
include Lada (1995), Bachiller (1996), Bachiller \& Tafalla (1999),
and Reipurth \& Bally (2001).  Massive YSO outflows are reviewed by
Garay \& Lizano (1999), Churchwell (1999), K\"onigl (1999), and
Beuther \& Shepherd (2005).  Richer et al. (2000), Cabrit (2002), and
Shepherd (2003) examine the similarities between massive and low-mass
outflows.  Here we examine new results associated with massive flows,
discuss a possible evolutionary scenario, and examine evidence for
morphological and dynamical differences between massive and low-mass
flows.

\vspace{-5mm}
\section{Outflow Energetics}\label{sec:energetics}

Young, low-luminosity YSOs ($L_{bol} \sim $ few~\Lsun) have mass
outflow rates $\dot M_f = 10^{-8}$ to $10^{-5}$~\Msunyr, momentum
rates $\dot P_f = 10^{-7}$ to $10^{-4}$~\Msunkmsyr, and mechanical
luminosity $L_{mech} = 10^{-3}$ to $10^{-1}$~\Lsun.  Young stars reach
the main sequence at about the same time the outflow terminates
($10^7$ to $10^8$~years).  The full opening angle, $\theta$, of flows
from low-luminosity YSOs tends to be 10--30\deg\ close to the central
source ($<$~50~AU) but then re-collimates ($\theta \sim$ few degrees)
within 100~AU from the protostar (e.g. Ray et al. 1996; Dougados et
al. 2000).
Intermediate-mass YSOs cover a wide range of $L_{bol}$.  For A-type
stars $L_{bol} \sim $ 10--30~\Lsun.  Outflows from A-type YSOs are
similar to those from low-mass YSOs although the energetics are
roughly an order of magnitude higher.  Well-collimated jets are common
and the associated molecular flows tend to be collimated for young
sources and less collimated for more evolved sources.  For B-type
stars $L_{bol}$ varies 2 orders of magnitude: from 30~\Lsun\ several
$\times 10^4$~\Lsun.  Early-B stars ($L_{bol} \sim 10^4$~\Lsun)
generate UC HII regions and reach the ZAMS while still accreting and
generating strong molecular outflows (e.g. Churchwell 1999, Garay \&
Lizano 1999, and references therein).  Intermediate-mass YSOs have
$\dot M_f = 10^{-5}$ to a few $\times 10^{-3}$~\Msunyr, $\dot P_f =
10^{-4}$ to $10^{-2}$~\Msunkmsyr, and $L_{mech} = 10^{-1}$ to
$10^2$~\Lsun.  Outflows from early-B and late O stars can be
well-collimated when the dynamical times scale is less than $\sim
10^4$ years although poorly collimated flows are more common in both
young and old sources.  
O stars with $L_{bol} > 10^4$~\Lsun\ generate powerful winds with
$\theta \sim 90$\deg\ within 50~AU of the star while accompanying
molecular flows can have $\theta > 90$\deg.  The flow momentum rate
($> 10^{-2}$~\Msunkmsyr) is more than an order of magnitude higher
than what can be produced by stellar winds and $L_{mech}$ exceeds
$10^2$~\Lsun\ (e.g. Churchwell 1999, Garay \& Lizano 1999).

For both well-collimated and poorly collimated flows, independent
studies have established correlations of the form: $\dot M \propto
L_{bol}^{0.6}$ where $\dot M$ is the bipolar molecular outflow rate
and the ionized mass outflow rate in the wind from $L_{bol} = 0.3$ to
$10^5$~\Lsun\ (e.g. Levreault 1988, Cabrit \& Bertout 1992, Shepherd
\& Churchwell 1996, Anglada 1996, Henning et al. 2000, Beuther et
al. 2002a, Wu et al. 2004).  There is also a strong correlation
between bolometric luminosity and circumstellar mass from $L_{bol} =
0.1$ to $10^5$~\Lsun\ (Saraceno et al. 1996, Chandler \& Richer 2000).
These correlations suggest that there is a strong link between
accretion \& outflow for a wide range of $L_{bol}$ - even into the
mid-O star range.

CO outflows from both low and high mass stars show a mass-velocity
relation in the form of a power law $dM(v)/dv \sim v^\gamma$ with
$\gamma$ ranging from $-1$ to $-8$.  The slope, $\gamma$, steepens
with age and energy in the flow (e.g. Rodr\'{\i}guez et al. 1982, Lada
\& Fich 1996, Shepherd et al. 1998, Richer et al. 2000).  A similar
relation of H$_2$ flux-velocity also exists with $\gamma$ between
$-1.8$ and $-2.6$ for low and high-mass outflows (Salas \&
Curz-Gonz\'alez 2002).

For a few early B (proto)stars with outflows that have a well-defined
jet, the jet appears to have adequate momentum to power the larger
scale CO flow although this relation is not as well established as it
is for lower luminosity sources.  For example, IRAS 20126$+$4104 has a
momentum rate in the SiO jet of $2 \times 10^{-1} \left( \frac{2
\times 10^{-9}}{{\rm SiO/H}_2} \right)$\,{\Msunkmsyr} while the CO
momentum rate is $6 \times 10^{-3}$\,{\Msunkmsyr} (Cesaroni et
al. 1999, Shepherd et al. 2000).  Although the calculated momentum
rate in the SiO jet is adequate to power the CO flow, the
uncertainties in the assumed SiO abundance makes this difficult to
prove.  Another example is IRAS\,18151--1208 in which the {\h} jet
appears to have adequate momentum to power the observed CO flow
(Beuther et al. 2002a, Davis et al. 2004).  For sources that have a
weak jet component coupled with a wide-angle flow, it has not been
determined what fraction of the momentum rate is supplied by the jet.
Despite the detection of jets toward massive protostars, the molecular
flows themselves tend to be less collimated than their low mass
counterparts.  Wu et al. (2004) find that the average collimation
factor (major/minor radius) for outflows from sources with $L_{bol} >
10^3$L{\sun} is 2.05 compared with 2.81 for flows from lower
luminosity sources.  This is true even for sources in which the
angular size of the flow is at least five times the resolution.  Thus,
the generally poorer collimation for massive flows is not due to low
resolution of the observations.

There are an increasing number sources for which our observations are
of comparable quality to those of low-mass YSOs and they provide a
quantitative view of the range of the outflow/infall properties for
these energetic sources.  Wu et al. (2004) provide a statistical
overview of the general properties of both massive and low-mass
outflows.  In the following sections, selected outflows from early B
and O (proto)stars are discussed and compared.

\vspace{-3mm}
\section{Well-Collimated Outflows from Early B (Proto)Stars}

Collimated, ionized jets can be generated by early-B protostars.  The
youngest sources ($\sim 10^4$ years or less) can be jet-dominated and
can have either well-collimated or poorly collimated molecular flows.
In a few sources, jets tend to have opening angles, $\theta$, between
25{\deg} and 30{\deg} but they do not re-collimate.  Other sources
appear to generate well-collimated jets ($\theta \sim$ few degrees)
that look like scaled up jets from low-luminosity protostars.  Jet
activity can continue as long as $10^6$ years although associated
molecular flows have large opening angles and complex morphology.
Below are some examples of sources with outflows that are at least
partially driven by a jet.

One of the best examples of an early B ZAMS star with a jet is
HH\,80-81 which has a highly collimated ionized jet with a projected
length $\sim 5$\,pc and age $\sim 10^6$\,years (Mart\'{\i},
Rodr\'{\i}guez, \& Reipurth 1993; Heathcote et al. 1998). The
truncated CO flow full opening angle is roughly 40\deg\ and does not
re-collimate (Yamashita et al. 1989).  The molecular flow momentum
rate ($\dot{P}_f = 6 \times 10^{-3}$~\Msunkmsyr) is an order of
magnitude greater than $\dot{P}_j$ in the ionized jet.  
The CO flow position angle is misaligned with the jet by roughly
30\deg.  The jet itself has only a slight wobble of a few degrees and
thus, jet precession is not likely to cause the wide opening in the
molecular flow.
The luminosity of the driving source is uncertain.  The total
luminosity of the system is $\sim 2 \times 10^4$\,L{\sun} however
there is at least one early B star near the jet center that could
account for a third or more of the luminosity (Stecklum et al. 1997).
Thus, HH\,80-81 appears to be powered by a B3-B1 ZAMS star.  A similar
situation is seen toward Ceph~A HW2 - an early B star with an ionized
jet, complex HH objects and multiple molecular outflows (e.g. Sargent
1979, Hartigan et al. 1986, Torrelles et al. 1993, Rodr\'{\i}guez et
al. 1994, and Garay et al. 1995).

While HH 80-81 looks like a scaled T-Tauri star with a jet that stays
well-collimated far from the central source, a growing number of
sources are being discovered which have jets with wider opening angles
that do not appear to re-collimate.  In particular, IRAS~20126$+$4104
(B0.5 spectral type, age $\sim $ few $\times 10^4$ years) has an
ionized and molecular jet which are misaligned from the larger scale
molecular outflow by more than 60\deg\ (Cesaroni et al. 1999, Cesaroni
et al. 2005, Hofner et al. 1999, Shepherd et al. 2000).  The large
precession angle in the IRAS\,20126 jet appears to be due to an
interaction with a close binary companion.  The full opening angle of
the $\sim$0.1\,pc SiO jet is roughly 40{\deg}.  This agrees with
recent estimates based on water maser studies that suggest the jet
full opening angle is $\sim 34${\deg} roughly 200 AU from the central
source (Moscadelli et al. 2005).

Another relatively wide-opening angle, ionized jet is
IRAS\,16547$-$4247 (Brooks et al. 2003, Garay et al. 2003,
Rodr\'{\i}guez et al. 2005).  The luminosity of the source suggests a
B0-O8 spectral type.  The molecular mass outflow rate is unknown but
the radio luminosity of the ionized jet is consistent with being
powered by a late O star.  Based on the extent of the {\h} emission,
the flow appears to be $\sim 7,000$ years old.  Rodr\'{\i}guez et
al. estimate the full opening angle of the jet to be $\sim 25${\deg}
out to about 10,000\,AU ($\sim 3''$ at D = 2.9 kpc).  Again, this is a
wide-opening angle jet relative to the typical 1-3{\deg} opening
angles often seen toward low-mass jets.  Two other examples of early-B
(proto)stars with well-collimated flows are: 1) IRAS\,05358$+$3543
which is an early B star cluster (age $\sim 4 \times 10^4$ years) that
has multiple collimated molecular outflows (Beuther et al. 2002b,
Sridharan et al. 2002); and 2) IRAS\,18151$-$1208 which harbors two
{\h} jets, one of which appears to be from an early B star (Davis et
al. 2004).

\vspace{-3mm}
\section{Wide-Angle Outflows from Early B (Proto)Stars}

Molecular outflows from early B (proto)stars are on average less
collimated than low-mass flows (Wu et al. 2004).  Low collimation factors
can occur in molecular flows even when there is a well-defined ionized
jet (see previous section).  In at least some sources, both the
ionized wind near the central source and the larger scale molecular
flow are poorly collimated and there is no evidence for a jet.
Several examples are given below.

The early-B YSO, G192.16--3.82, has a poorly collimated ionized wind
($\theta \sim 40$\deg) within 50~AU of the YSO that expands to $\theta
\sim 90$\deg\ 0.1~pc from the source (Shepherd et al. 1998, Shepherd
\& Kurtz 1999, Shepherd, Claussen, \& Kurtz 2001).  The 100\,M{\sun}
molecular flow is a few $\times 10^5$\,years old and forms the
truncated base of the larger scale flow that extends more than 10~pc
from end-to-end (Devine et al. 1999).  The outflow is consistent with
being produced by a wind blown bubble and there is no evidence for a
collimated jet.  A source that appears to be similar to G192.16 is
AFGL 490.  It has a wide angle wind and outflow from an early B star
(e.g. Snell et al. 1984, Mitchell et al. 1995, Schreyer et al. 2005).
The age of the CO flow is estimated to be $\sim 2 \times 10^4$ years
old, the outflow opening angle is roughly 50{\deg} and the collimation
factor is about 1.0.  The central star, detected in the near-IR
(e.g. Davis et al. 1998) is surrounded by a 20,000 AU CS torus that
appears to have an inner accretion disk that is less than 500 AU in
diameter (Schreyer et al. 2002).  There is no obvious sign of a
collimated jet from the central object.

As a final example consider W75\,N.  At the center of the CO outflows
is a cluster of UC HII regions.  One, VLA\,2, appears to power a
wide-angle, red-shifted CO flow to the south-west (Davis et al. 1998,
Shepherd 2003, Shepherd et al. 2004).  Proper motions of water masers
associated with the outflow from VLA\,2 show that the opening angle of
the flow appears to be nearly 180{\deg} within 100\,AU of the star
(Torrelles et al. 2003).  The larger scale CO flow opening angle is
roughly 50{\deg}, consistent with a wind-blown bubble.  There is no
indication near the star that a well-collimated jet exists.  The
infrared line emission suggests that only slow, non-dissociative
J-type shocks exist throughout the parsec-scale outflow.  Fast,
dissociative shocks, common in jet-driven outflows from low-mass
stars, are absent in W75\,N.  Interestingly, VLA\,1, just 2,000 AU
from VLA\,2, is a jet that drives a collimated CO flow.  Torrelles et
al. (2004) suggest that the increased collimation may be because
VLA\,1 is at a different evolutionary state than VLA\,2.

\vspace{-4mm} 
\section{Mid to Early O Star Outflows}

To date, extremely collimated outflows have not been observed toward
sources earlier than B0.  It is possible that this is simply a
selection effect because O stars form in dense clusters and reach the
ZAMS in only a few $\times 10^4$ years.  Thus, any collimated outflows
may be confused by other flows.  For a few sources that are relatively
isolated and for which adequate high-resolution data are available, no
inner accretion disk ($\sim 50$~AU diameter) has been conclusively
detected. In a few sources, collimation close to the protostar appears
to be due to pressure confinement from an equatorial torus of dense
gas (e.g. K3-50A: Depree et al. 1994, Howard et al. 1997).  This may
also be the case for the Orion\,I outflow (e.g. McCaughrean \& Mac Low
1997, Greenhill et al. 1998) however, there is now strong evidence
that Source I and the BN object are moving away from each other and
that they were within 225 AU of each other about 500 years ago
(Rodr\'{\i}guez et al. 2005).  Source I and BN may have been members
of a multiple system or BN may have been a companion of
$\theta^1$\,Ori\,C and made a close approach to Source I after it was
ejected (Tan 2003).  The poorly collimated, explosive-looking outflow
seems to have occurred $\sim 1,000$ years ago.  Could it have been
caused by the event that ejected the BN object?  If so, then the
outflow is not actively powered by Source I and the geometry of the
circumstellar material around Source I need not be linked to the
large-scale flow.  Such explosive events may not be uncommon as
evidenced by the 'fingers' of shocked gas seen in Spitzer images of
the ionized outflow associated with G34.26$+$0.15 (Churchwell,
personal communication).

As a final example, consider G5.89--0.39.  The UC HII region appears
to be powered by an O5 star.  The star has a small excess at 3.5$\mu$m
suggesting the presence of circumstellar material (Feldt et al. 2003).
The O5 star is along the axis of two {\h} knots that appear to trace a
collimated N-S flow (Puga et al., poster presented at this
conference).  There is also a N-S C$^{34}$S(J=3-2) and the UC HII
region is expanding in the N-S direction (Cesaroni et al. 1991, Acord
et al. 1998).  Although still circumstantial, the evidence is mounting
that the O5 star in G5.89 produced the N-S outflow and thus is forming
via accretion.  Note, there is also an SiO flow (with a NE-SW
orientation) that is powered by a star to the S-W of the O5 star
(Sollins et al. 2004, Puga et al. in preparation) and a larger scale
E-W CO flow for which the driving source has not been identified
(Watson et al. 2002).

Additional evidence that O stars form via linked accretion and outflow
comes from 7\,mm continuum observations of very young O stars (van der
Tak \& Menten 2005).  Models of the derived sizes, flux densities, and
radio spectra of sources with luminosities up to $\sim 10^5$L{\sun}
suggest that at least one star has a dust disk and all are accreting
material.  Although not conclusive, this supports the accretion-based
formation scenario for even mid-O stars.

\vspace{-3mm}
\section{Evidence for an Evolutionary Scenario}\label{sec:evolution}

Current observations indicate that the outflow/infall mechanism is
similar from T~Tauri stars up to early B protostars.  Although there
is evidence that the energetics for at least some early-B stars may
differ from their low-mass counterparts, the dynamics are still
governed by the presence of linked accretion and outflow.  A few young
O stars show evidence for accretion as well.  Although there are
clearly explosive events, these may be due to close encounters that
disrupt the accretion process.  However, dynamical mergers of
(proto)stars to create the most massive stars may still be a viable
formation mechanism.


There are both well-collimated and poorly collimated molecular flows
from massive stars.  Well-collimated molecular flows tend to be in
systems with ages less than a few times $10^4$ years old where the
central object has not yet reached the main sequence.  Hence the
effects of increased irradiation on the disk and disk-wind due to the
stellar radiation field are minimized.  Observed jets often have
opening angles between 25{\deg} and 30{\deg} with little evidence for
recollimation of the jets on larger scales.  This could be due to a
change in the balance between magnetic and plasma pressure.  Poorly
collimated flows (opening angle greater than 50{\deg} that show
no evidence for a more collimated component) are associated with more
evolved sources that have detectable UC HII regions and the central
star has reached the main sequence.  Thus, the disk and outflow are
subject to significantly more ionizing radiation.

To account for the differences seen in flow morphologies from early B
to late O stars Beuther \& Shepherd (2005) proposed two
possible evolutionary sequences which could result in similar
observable outflow signatures as shown in Figure 1.
For both early B and O stars, the flows begin collimated and become
less-collimated as the star reaches the ZAMS and generates
significantly more Lyman continuum photons.  This evolutionary
sequence appears to qualitatively fit the observations however it must
be tested against both theory and observations.

\subsection{Impact of Luminosity on Outflows}

Once a massive OB star reaches the main sequence, the increased
radiation from the central star generates significant Lyman continuum
photons and will likely ionize the outflowing gas even at large radii.
The result may be an increase in the plasma pressure at the base of
the flow which could overwhelm any collimating effects of a magnetic
field (see, e.g., Shepherd 2003, K\"onigl 1999).  It may also affect
the inherent collimation of the ionized wind from the {\it stellar}
surface as suggested in the simulations by Yorke \& Sonnhalter (2002).
Recollimation of the outflowing gas from the disk is expected under
ideal magneto-hydrodynamic (MHD) conditions where the recombination
timescale is much greater than the dynamical timescale of the outflow.
Thus, the ionized plasma becomes 'frozen-in' to the magnetic field
lines.  Models of the ionization \& density along beams of HH jets
indicate that ideal MHD is valid for partially ionized jets from T
Tauri stars (e.g. Bacciotti \& Eisl\"offel 1999).  However, the ideal
MHD assumption breaks down as: 1) the plasma temperature and density
increase; 2) the turbulence in the disk or wind increases; or 3) the
toroidal component of the magnetic field, $B_{\phi}$, decreases.  As
$L_{bol}$ increases plasma temperature and density will increase and
the disk may be more turbulent.  Thus, ideal MHD assumptions may break
down for high-mass outflows and one may expect that energetics and
re-collimation could be affected.  To be more specific, conditions
appropriate for luminous YSO must be included in simulations
(e.g. K\"onigl 1999).  It is interesting to note that several early
B protostars with strong jet components in the outflows have $\sim
30${\deg} opening angle jets that do not appear to re-collimate.
Perhaps this is due to a break-down in ideal MHD conditions along the
outflow cone?

\subsection{Impact of Mass-Loading on Outflows}

Two conditions expected for luminous YSOs are increased disk surface
heating (due to increased stellar radiation and shocks) and a higher
level of disk turbulence when $M_D/M_{\star} > 0.3$ where $M_D$ is the
disk mass and $M_{\star}$ is the (proto)stellar mass.  Increased disk
surface heating causes increased mass loading (e.g. Ferriera 2002)
and since winds carry angular momentum away, this means that a warmer,
denser wind causes the disk to rotate slower.  Pudritz (these
proceedings) has evaluated the impact of mass loading on magnetized
disk-winds under conditions appropriate for massive protostars that
have not yet reached the ZAMS (e.g. no significant heating from the
central protostar).  Solutions show that a strong wide-angle wind
develops as the mass-loading increases.  Thus, the outflow dynamics
and geometry appears to change as the outflow rate increases.

\subsection{Impact of Disk Turbulence on Outflows}

Disks around early-B YSOs are significantly more massive than those
around A-type and T~Tauri YSOs.  Further, $M_D/M_\star$ is often
greater than 0.3 which may cause the disk to be locally unstable and
provide an additional means of angular momentum transport from the
protostar (e.g. Laughlin \& Bodenheimer 1994; Yorke, Bodenheimer, \&
Laughlin 1995; Shepherd, Claussen, \& Kurtz 2001).  There are a
growing number of early-B (proto)stars that are actively powering
molecular outflows and have circumstellar disks or molecular tori with
masses ranging from 0.1\,{\Msun} to tens of solar masses.  
The increased $M_D/M_\star$ suggests that the detailed dynamics of
infall and outflow in early-B YSOs may differ from their lower mass
counterparts.  Recently, Lodato \& Rice (2005) have
made simulations of circumstellar disks in which $M_{disk}$ = 0.1,
0.5, and 1.0\,M${_\star}$ to evaluate the behavior of spiral density
waves.  They find that the mass of the disk dramatically changes the
characteristics of the spiral density waves that can transport angular
momentum efficiently.  As the disk mass increases, the waves become
stronger and in some cases episodic.  The associated accretion onto
the central star increases during periods when spiral density waves
are strong and carry away significant angular momentum.  It is not
clear how or even if such behavior will cause significant changes in
the outflow characteristics but one might expect it would since
accretion and outflow are so inherently linked.  The simulations of
Lodato \& Rice are in a regime in which fragmentation of the disk is
not allowed, further, only disk heating and cooling are considered (no
stellar heating or heating due to the outflow or shocks above the disk
plane).  Additional improvements to the simulations are planned which
will incorporate more realistic heating and cooling functions and disk
heating from the star and outflow.  Indeed, simulations of disks
around low-mass stars which include realistic heating from the central
star and outflow suggest that irradiation can quench fragmentation due
to local gravitational instabilities because the disk temperature is
raised above the parent cloud temperature (Matzner \& Levin 2005).

The simulations of Lodato \& Rice (2005) suggest an increase in ${\dot
M_{acc}}$ if spiral density waves transport angular momentum to the
outer parts of the disk - one would not expect a corresponding
increase in the disk wind mass loss rate, ${\dot M_w}$, since there is
not excess angular momentum in the inner disk where the wind is
generated.  Thus, significant angular momentum transport associated
with spiral density waves in the disk may contribute to a decrease in
${\dot M_w}/{\dot M_{acc}}$ as well.  Is there observational evidence
for a decrease in ${\dot M_w}/{\dot M_{acc}}$?  Richer et al. (2000)
demonstrate that one can measure the Outflow force/Accretion Force vs
$L_{bol}$ and that this quantity can be related to ${\dot M_w}/{\dot
M_{acc}}$.  Shepherd (2003) shows that there is marginal evidence for
such a decrease but the errors in the data are large and any trend is
at best, marginal.  Further observations are necessary to determine if
such a trend actually exists.

\vspace{-4mm}
\section{Summary}\label{sec:summary}
Mid- to early-B protostars and late O protostars have accretion disks
and outflows that can be well-collimated.  Ionized jets in very young
sources (age {\ltsim}\,$10^4$ years) can have opening angles $\sim
30${\deg} with no recollimation detected or they can re-collimate
similar to jets from low-luminosity protostars.  Once an HII region
forms, the associated ionized outflows have strong wide-angle winds
and a collimated jet is often not detected.  The non-detection of a
jet in older sources is unlike low-mass outflows were there is
evidence for a two-component wind (jet$+$wide-angle) even in older
sources {e.g. Arce \& Goodman 2002).  Mid to early O stars have
poorly collimated outflows and explosive events suggest that
protostellar or protostar/disk interactions are common.  Current
observations suggests that outflows from early OB protostars begin
collimated and then become less collimated with time once the star
reaches the main sequence.  This evolution in outflow dynamics and
morphology could be due to, e.g.: increased ionizing radiation from
the star as it begins to burn hydrogen; mass-loading onto the disk
wind; and/or disk instabilities which can be an efficient angular
momentum transport mechanism in massive disks.


\clearpage

\begin{figure}
\epsscale{1}
\plotone{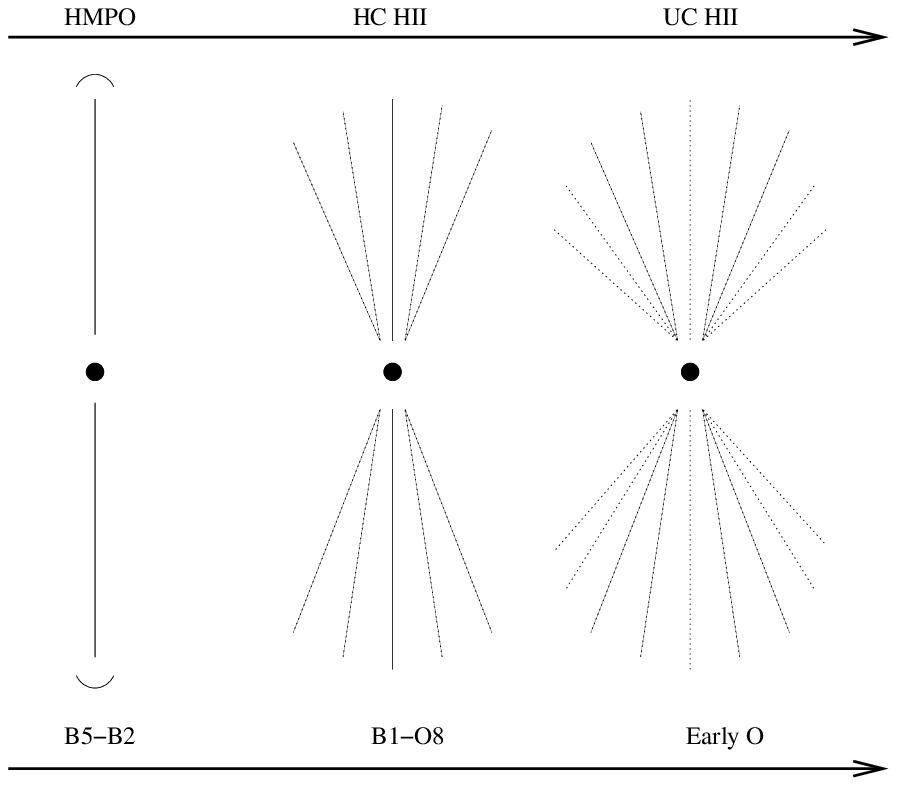}

\vspace{5mm}
\caption{Sketch of the proposed evolutionary outflow
scenario (\cite{bs05}).  The three outflow morphologies can be
caused by two evolutionary sequences: (top) the evolution of a typical
B1-type star from a high-mass protostellar object (HMPO) via a
hyper-compact H{\sc ii} (HC\,HII) region to an ultra-compact H{\sc ii}
(UC\,HII) region, and (bottom) the evolution of an O5-type star which
goes through B1- and O8-type stages (only approximate labels) before
reaching its final mass and stellar luminosity.}
\end{figure}

\end{document}